# Theoretical single droplet model for particle formation in flame spray pyrolysis

Yihua Ren[*], Jinzhi Cai, Heinz Pitsch

Institute for Combustion Technology, RWTH Aachen University, Templergraben 64, Aachen 52056, Germany

[*]Corresponding author: Yihua Ren

Mail Address: Institute for Combustion Technology, RWTH Aachen University, Templergraben 64, Aachen 52056, Germany

*Email:* y.ren@itv.rwth-aachen.de

**Colloquium:** SOOT, NANOMATERIALS, AND LARGE MOLECULES including the formation, growth, and destruction of soot, PAHs, carbon nanostructures, and other nanoscale materials

Total Length:   5740 words, Method 1

Main Text:  2964 words (MSWord 2013 word count)

Equations 1: (1 equation lines + 2 blank lines) x (7.6 words/mm) x (1 column) = 22.8 words

Equations 2: (1 equation lines + 2 blank lines) x (7.6 words/mm) x (1 column) = 22.8 words

Equations 3: (2 equation lines + 2 blank lines) x (7.6 words/mm) x (2 column) = 60.8 words

Equations 4: (1 equation lines + 2 blank lines) x (7.6 words/mm) x (2 column) = 45.6 words

Equations 5: (1 equation lines + 2 blank lines) x (7.6 words/mm) x (2 column) = 45.6 words

Equations 6: (2 equation lines + 2 blank lines) x (7.6 words/mm) x (2 column) = 60.8 words

Equations 7: (1 equation lines + 2 blank lines) x (7.6 words/mm) x (2 column) = 45.6 words

References:   (21 references + 2) x (2.3 lines/reference) x (7.6 words/line) = 402 words

Figure 1: (109.73mm+10mm) x 2.2 words/mm x 1 column + 41 words in caption = 304 words

Figure 2: (67.22  mm+10mm) x 2.2 words/mm x 1 column + 9   words in caption = 179 words

Figure 3: (81.83  mm+10mm) x 2.2 words/mm x 1 column + 56 words in caption = 258 words

Figure 4: (61.98  mm+10mm) x 2.2 words/mm x 2 column + 91 words in caption = 407 words

Figure 5: (50.91  mm+10mm) x 2.2 words/mm x 1 column + 25 words in caption = 159 words

Figure 6: (50.80  mm+10mm) x 2.2 words/mm x 1 column + 17 words in caption = 151 words

Figure 7: (43.97  mm+10mm) x 2.2 words/mm x 2 column + 42 words in caption = 279 words

Figure 8: [104.88mm+10mm] x 2.2 words/mm x 2 column + 63 words in caption = 568 words

Figure 9: [104.88mm+10mm] x 2.2 words/mm x 2 column + 63 words in caption = 568 words

**Abstract**

In the current work, we develop a single droplet model to describe particle formation in multicomponent-liquid droplet combustion. Both the gas-to-particle conversion and droplet-to-particle conversion routes of different solution properties are investigated, together with the population balance model and the droplet drying model. For multi-component droplets without precursors, the droplet combustion is limited by the species diffusion in the liquid phase. The model can well predict the droplet shrinkage history observed by previous PDA measurements. For the precursor with a low boiling point than its thermal decomposition temperature, the precursor in the droplet can then transform into nanoparticles through the gas-to-particle conversion route. The population balance model reveals that the generated nanoparticle size relies on both the precursor mass fraction and the residence time, which is consistent with the vapor-fed aerosol flame synthesis. For the precursor that tends to decompose or precipitate in the liquid, it then undergoes the droplet-to-particle conversion route. The droplet behavior can be classified by the ratio of droplet evaporation time and precursor reaction or precipitation time. For small droplets with short evaporation time, the nanoparticle formation obeys the one-droplet-one-particle rule. For large droplets with long evaporation time, the competition among precipitation, thermal decomposition, and evaporation determines the final nanoparticle morphology. The single droplet model established in this study can potentially guide the precursor design and be coupled with the turbulent flame simulation of the whole flame spray pyrolysis burner.

2

# 1. Introduction

Flame spray pyrolysis (FSP) has emerged as a promising technology for industrial production of single- and multi-component metal oxide nanomaterials with a high yield of $10^2$-$10^3$ g/h [1–3], with notable applications including catalysts [4], optical materials [5], sensors [6], energy storage materials [7,8], *etc*. Different from conventional vapor-fed aerosol flame synthesis (VAFS) method [9,10], FSP can use less volatile precursors like nitrates, which significantly lowers the cost and extends the versatility in elements. In FSP, the precursor dissolved in combustible solvents can be atomized and ignited to generate a self-sustaining spray flame. Starting from the droplet evaporation, nanoparticles are generated through a series of physio-chemical processes in turbulent spray flames. Numerical simulation of such a combustion system is quite challenging due to the turbulent flame coupled with complex phase transitions and homo- and heterogeneous reactions during the conversion from droplets to nanoparticles. A theoretical single droplet model that bridges the microscopic droplet reaction and the macroscopic turbulent combustion can significantly benefit the numerical simulation of the whole FSP process, guide the design of the liquid precursors, and optimize FSP reactors.

Depending on the solution property and temperature-time history, the transformation processes can be divided into the gas-to-particle conversion and the droplet-to-particle conversion routes. For the gas-to-particle conversion, volatile precursors transform to gases and transfer into nanoparticles through reaction, nucleation, collision, coalescence (or sintering), which is similar to reactions in VAFS and soot formation. In contrast, the droplet-to-particle conversion corresponds to the one-droplet-one-particle route, as the non-volatile precursors precipitate and decompose inside the droplets, which gives rise to sub-micron or hollow particles. Jossen *et al* [11] proposed two criteria to characterize the conversion routes in FSP, i.e. (1) the ratio of the solvent boiling temperature over the precursor decomposition temperature, (2) the combustion enthalpy density. Strobel and Pratsinis [12] later investigated the effects of solvent composition and found that the additive liquid 2-ethylhexanoic acid (2-EHA) could significantly enhance the gas-to-particle conversion routes compared with pure ethanol as solvent. Wei *et al* [13] examined the droplet behavior with different solvent compositions and considered the formation of low-boiling-point 2-ethylhexanates from nitrates due to the ligand exchange effect. Mädler and co-workers [14,15] proposed a micro-explosion mechanism of multicomponent-liquid droplets by observing the combustion of an isolated 100 μm droplet using rainbow refractometry. Abram *et al* [16] emphasized the effect of droplet size on the particle morphology and built a qualitative map of particle formation route. Therefore, a quantitative model accounting for various factors in the particle formation routes is urgently needed to facilitate precursor design, numerical modeling, and active control of the FSP.

From the computational aspect, a multicomponent single-droplet combustion model has been widely utilized in analyzing liquid fuel combustion [17,18]. Depending on the Lewis number of species, droplet combustion can be divided into the well-mixed regime and the diffusion-limited regime. Gröhn *et al.* [19] simulated volatile precursor droplets by assuming well-mixed liquid and coupled it with the continuum gas equation of a spray flame. Li *et al.* [15] established a diffusion-limited single droplet model considering concentration gradients inside the droplet. However, fewer efforts were devoted to modeling the transformation from droplets to nanoparticles through conversion routes considering the gas, droplet and liquid phases.

In this work, we propose a multi-component liquid droplet combustion and particle formation model considering both the gas-to-particle conversion and droplet-to-particle conversion routes with different solution properties. For the gas-to-particle conversion, a population balance model is further solved for the duration of the droplet lifetime. For the droplet-to-particle conversion, a droplet drying procedure is simulated and investigated to predict the particle morphology.

## 2. Theoretical Model

The single droplet model describes the reaction and transport in the gas, liquid, and particle phases. The algorithm map of the single droplet model is shown in Fig. 1. Based on the temperature and mass fraction at the previous time step from the liquid phase, the temperature and mass fraction solution of the gas phase are directly calculated. The temperature and species gradients are then utilized as boundaries conditions for the multi-component droplets. Depending on whether the liquid precursor reacts in the liquid phase or evaporates into the gas phase, the gas-to-particle or droplet-to-particle conversion route is solved based on the population balance model or the droplet drying model, respectively. Finally, as the droplet shrinks, we move to the next time step and consider the gas phase transport.

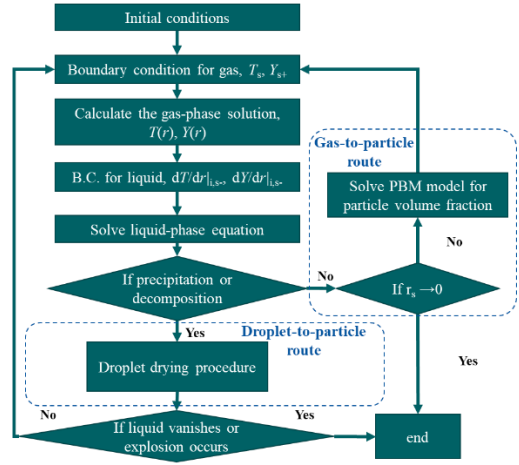

Fig. 1. The algorithm map of the single droplet model

**2.1 Gas phase analysis**

For the gas phase, the spherically-symmetric vaporization and combustion of a multi-component droplet are modeled with an initial radius of $r_{s0}$ and temperature of $T_0$ in the ambient pressure $P_0$. At infinity, the atmosphere has the oxygen mass fraction $Y_{O\infty}$ and the temperature $T_\infty$. The species mass fraction at the droplet surface is calculated according to the Raoult's Law,

$$x_{i,s^+} = x_{i,s^-} \exp(\frac{L_{vi}}{R}(\frac{1}{T_{bi}} - \frac{1}{T_s})), \tag{1}$$

where $x_{i,s^-}$, $x_{i,s^+}$ represent molar components fraction in droplet surface of liquid and gas phase, respectively, $L_{vi}$ is the latent heat of species $i$, $T_{bi}$ is the boiling point, and $T_s$ is the surface temperature. Here, both $T_s$ and $x_{i,s^+}$ are obtained from the solutions of the liquid-phase equation in the previous step. For simplicity, we invoke the assumptions of Le = 1 for all gaseous species and consider the Burke-Schumann limit that the infinitely-thin flame sheet stays at $r_f$ in a quasi-steady manner. Following previous works, the Spalding mass transfer number $B_m$ can be expressed as

$$B_m = \frac{Y_{O\infty}/\nu + \sum_i Y_{Fi,s^+}}{1 - \sum_i Y_{Fi,s^+}}, \qquad (2)$$

where $\nu$ represents the stoichiometric number, $Y_{Fi,s^+}$ is the mass fraction of the fuel species $i$ in the gas phase at the droplet surface. Based on the relation, we can derive the droplet evaporation rate as $\dot{m} = 4\pi r_s \rho_g D_g \ln(1 + B_m)$ where $\rho_g$ and $D_g$ are the gas density and gas diffusivity. Profiles of the temperature $T(r)$ and species mass fraction $Y_{Fi}(r)$ can be then calculated by integrating conserved scalars. A typical flame sheet structure is demonstrated in Fig. 2, the fuel and oxidizer react at the spherical diffusion flame with the flame sheet locating at the radial position $r_f$.

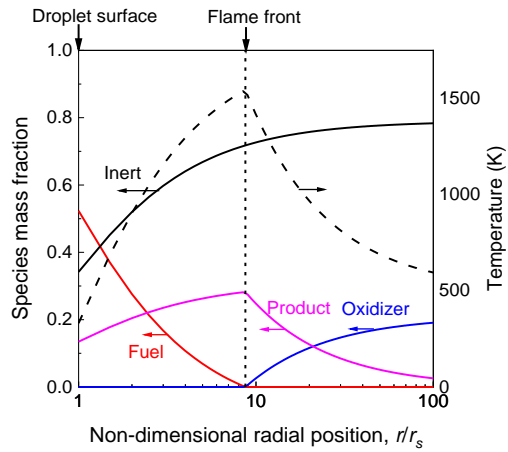

Fig. 2. Simulated distributions of the mass fractions of fuel, oxidizer, products and inert species, as well as the temperature profile based on the single droplet model. $T_\infty = 500$ K, $Y_{O\infty} = 0.21$, $r_s^2/r_{s0}^2 = 0.99$, $Y_{\text{ethanol}} = 0.9$, $Y_{\text{2-EHA}} = 0.1$.

## 2.2 Liquid-phase analysis

The mass and heat transport of the droplet liquid are modeled in a spherically symmetric form. Considering the coordinate transformation $R=r/r_s$, the model can be transformed into a fix-boundary problem which can be expressed as,

$$\begin{cases} \dfrac{\partial T_l}{\partial t} - \dfrac{\dot{r}_s}{r_s} \cdot R \dfrac{\partial T_l}{\partial R} = \dfrac{\alpha_l}{r_s^2} \cdot \dfrac{1}{R^2} \dfrac{\partial}{\partial R}\left(R^2 \dfrac{\partial T_l}{\partial R}\right) \\ \dfrac{\partial Y_{i,l}}{\partial t} - \dfrac{\dot{r}_s}{r_s} \cdot R \dfrac{\partial Y_{i,l}}{\partial R} = \dfrac{\alpha_l}{r_s^2 \text{Le}_{i,l}} \cdot \dfrac{1}{R^2} \dfrac{\partial}{\partial R}\left(R^2 \dfrac{\partial Y_{i,l}}{\partial R}\right) \end{cases} \quad (3)$$

where $T_l$ is the liquid-phase temperature, $Y_{i,l}$ is the mass fraction of species $i$, $\alpha_l$ is the thermal diffusivity, $\text{Le}_{i,l}$ are the Lewis numbers of the species $i$. In this equation, the second term on the left-hand-side is an additional convection term due to the shrinking droplet with the speed of $\dot{r}_s$.

Besides, the mass fractions and temperatures across the liquid-gas interface obey the balance relation,

$$\dot{m} Y_{il,s^-} - 4\pi r_s^2 \rho_l D_{l,i} \left.\dfrac{\partial Y_{il}}{\partial r}\right|_{r_s^-} = \dot{m} Y_{Fi,s^+} - 4\pi r_s^2 \rho_g D_g \left.\dfrac{\partial Y_{Fi,s^-}}{\partial r}\right|_{r_s^-} \quad (4)$$

$$4\pi r_s^2 \lambda_l \left.\dfrac{\partial T_l}{\partial r}\right|_{r_s^-} + \dot{m} \sum_i L_v Y_{il,s^-} = 4\pi r_s^2 \lambda_g \left.\dfrac{\partial T}{\partial r}\right|_{r_s^+} \quad (5)$$

where $\lambda_g$ and $\lambda_l$ are the heat conductivity of the gas phase and liquid phase, respectively, the subscript $s^-$ and $s^+$ represent the liquid-phase and gas-phase conditions at the gas-liquid interface. Equations (4) and (5) give the boundary condition at the droplet surface, while at the central position, the species mass fraction and temperature have zero gradients.

If the precursor does not evaporate but precipitates or decomposes in the liquid phase, the droplet will undergo a drying procedure. In this condition, this non-volatile species has a Stefan-flow condition at the gas-liquid interface, as the right-hand side in Eq. (4) is zero. In addition, an additional source term is added onto the right-hand side of Eq. (3) to describe the precipitation of the non-volatile species in the liquid phase. To simplify our analysis later, we consider the nucleation rate is equal to $k(Y_{il}-Y_{il}^*)^3$ where $Y_{il}^*$ is the critical mass fraction for precipitation. The characteristic precipitation time scale is estimated by $1/k$.

## 2.3 Particle-phase analysis

If the liquid precursor can be vaporized, we consider that the precursor is decomposed in the gas phase and transforms into solid monomers. The nucleation is assumed to happen instantaneously because, for most metal oxides, the critical nucleation size is smaller than the monomer size. Then, the convection, diffusion, and coagulation of nanoparticles are described by a convection-diffusion equation coupled with a monodisperse population balance model in the region $R > 1$:

$$\begin{cases} \dfrac{\partial \phi}{\partial t} - \dfrac{\dot{r}_s}{r_s} R \dfrac{\partial \phi}{\partial R} + \dfrac{1}{r_s^2} \dfrac{1}{R^2} \dfrac{\partial}{\partial R}\left[ R^2 \left( (u+c) r_s \phi - D_p \dfrac{\partial \phi}{\partial R} \right) \right] = S_p \\ \dfrac{\partial n}{\partial t} - \dfrac{\dot{r}_s}{r_s} R \dfrac{\partial n}{\partial R} + \dfrac{1}{r_s^2} \dfrac{1}{R^2} \dfrac{\partial}{\partial R}\left[ R^2 \left( (u+c) r_s n - D_p \dfrac{\partial n}{\partial R} \right) \right] = \dfrac{S_p}{v_{p0}} - \dfrac{1}{2} \beta n^2 \end{cases} \quad (6)$$

where $\phi$ is particle volume fraction, $n$ is the particle number density, $u$ is the flow velocity obtained from the gas-phase equation, $c$ is the thermophoretic velocity, $D_p$ is the nanoparticle Brownian diffusion, $v_{p0}$ is the monomer size, $\beta$ is the Brownian coagulation rate. The expressions of $D_p$, $c$, and $\beta$ can be found in Ref. [20]. The source term $S_p$ is related to the vapor precursor reaction at the flame front, which can be expressed as

$$S_p = \left( \dot{m} Y_{Fp,s^+} - 4\pi r_s^2 \rho_g D_g \dfrac{\partial Y_{Fp}}{\partial r}\bigg|_{r=r_s^+} \right) \dfrac{\delta(R - r_f/r_s)}{4\pi r_f^2 r_s} \dfrac{v_p}{\rho_{particle}}, \quad (7)$$

where $\delta$ is a non-dimensional delta function, $\rho_{particle}$ is the particle density, the subscript $p$ represents the precursor, and $v_p$ is the stoichiometric factor of the reaction from precursor to nanoparticles, defined as $\nu MW_m/MW_p$.

## 2.4 Numerical method

The single droplet model defined by Eqs. (1)-(7) with the corresponding boundary conditions is solved numerically to compute the profiles of gas-phase, liquid-phase, and particle-phase. The liquid-phase equations are discretized by the central difference scheme and solved by the implicit-Euler method and the LU decomposition of the sparse matrix in MATLAB software. The particle-phase equations are discretized by a first-order upwind scheme and solved explicitly by the Runge-Kutta method using ode45. The grid convergence was verified with consecutively refined grids.

## 3. Result and Discussion

### 3.1 Multicomponent droplet combustion

The combustion process of a multi-component droplet in absence of nanoparticle precursor is investigated first. Here we consider a mixture of ethanol and 2-EHA which is widely utilized for FSP. The droplet shrinking process is shown in Fig. 3. The surface temperature is not completely consistent with the central temperature, as the droplet thermal diffusion time scale, $r_s^2/\alpha_L \sim 2.8$ μs is comparable with the droplet lifetime $r_{s0}^2/(\rho_g D_g \cdot \ln(1+K)) \sim 7.8$ μs, especially at the early stage of droplet shrinking. Moreover, since liquid-phase mass transfer is much slower than the thermal transfer, the droplet gasification process includes fast vaporization of the volatile species in the beginning (ethanol with boiling point of 355 K) and then accumulation of less volatile species (2-Ethylhexanoic acid, 2-EHA, with boiling point of 487 K) at the outer layer accordingly. When the droplet shrinks to a small scale, flame extinction happens and the droplet vaporizes without combustion. By comparing with the PDA measurements based on Ref. [13], flame extinction occurs when the square of the normalized droplet size, $r_s^2/r_{s0}^2$, reaches 0.2.

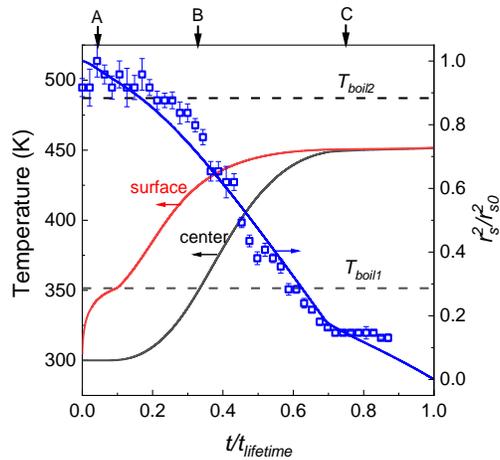

Fig. 3. Temporal variation of surface and center temperatures of a multi-component droplet as the droplet shrinks in a $D^2$-law, which is compared with the PDA measurement data (blue squares) based on Ref. [13]. $T_\infty = 1000$ K, $Y_{O\infty} = 0.21$, $Y_{\text{ethanol}} = 0.5$, $Y_{\text{2-EHA}} = 0.5$, $\text{Le}_l = 10$.

Figure 4 further shows the spatial distributions of temperature and species mass fraction at different moments labeled as A, B, and C in Fig. 3. In the beginning, the volatile ethanol evaporation dominates, while 2-EHA stays at the inner layer. Then, as the surface temperature gradually increases to the boiling point of ethanol, the ethanol evaporates quickly so that the 2-EHA concentrates at the surface region. After the temperature exceeds the boiling point of ethanol, the ethanol mole fraction in the gas phase is equal to its mole fraction in the liquid phase. As the volatile species is trapped inside the droplet, the superheat or internal boiling of the liquid could occur. Previous investigations [17] indicate that the internal bubbling of the superheated liquid due to homogeneous nucleation is difficult to achieve at the atmospheric ambient pressures, while the internal boiling due to heterogeneous nucleation makes it much easier to reach. For flame spray pyrolysis, this could be induced by precipitation or thermal decomposition reaction during the droplet combustion, as observed in the single droplet combustion experiment by Li *et al.* [15].

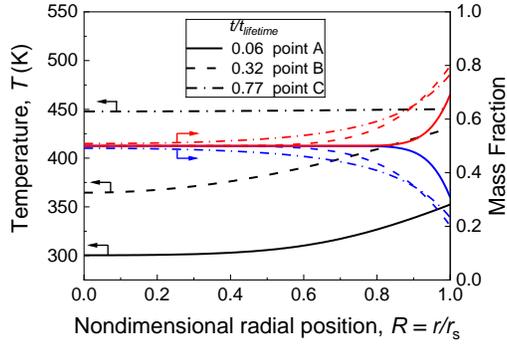

Fig. 4. Spatial distributions of temperature and species mass fractions of ethanol (blue) and 2-EHA (red) at different moments, labeled as A, B, C in Fig. 3. $T_\infty = 1000$ K, $Y_{O\infty} = 0.21$, $Y_{ethanol} = 0.5$, $Y_{2\text{-EHA}} = 0.5$.

**3.2 Gas-to-particle conversion route**

When the liquid precursor can be vaporized, it transforms into nanoparticles through the gas-to-particle conversion routes. Depending on the precursor reaction mechanism, the precursor will react between the droplet surface and the flame sheet. In this work, we focus on the particle formation with the precursor reaction at the flame sheet, as shown in Fig. 5. Two different situations are analyzed. For the first one, the metal-organic precursor with its boiling point $T_{bp,precursor}$ higher than the solvent boiling point $T_{bp,solvent}$ is simulated. Here we choose the thermophysical properties of 2-EHA as characteristic properties of the precursor, as referred to Ref [15]. A significant sign is the flame shrinkage phenomenon, as displayed by the flame sheet position. This phenomenon is attributed to the variation of the evaporation rate after the quick vaporization of the volatile solvent at the surface [21]. Followed by the flame resumption, the particle volume fraction starts to grow as the precursor species dominates the vaporization of the droplet. The primary nanoparticles that form at the flame sheet coagulate and then are transported to the droplet surface by thermophoresis and diffusion, and towards infinity. On the other hand, if $T_{bp,precursor}$ is lower than $T_{bp,solvent}$, the precursor vaporizes in the beginning, and the nanoparticles are generated during the whole lifetime of the droplet. In both conditions, the low radial flow velocity leads to the accumulation of nanoparticles so that the nanoparticles coagulate outside the flame sheet. In this region, however, the temperature-time history, the number density, and the collision kernel of nanoparticles could be primarily influenced by the external flow field, which we need to be considered in the turbulent combustion model in the future.

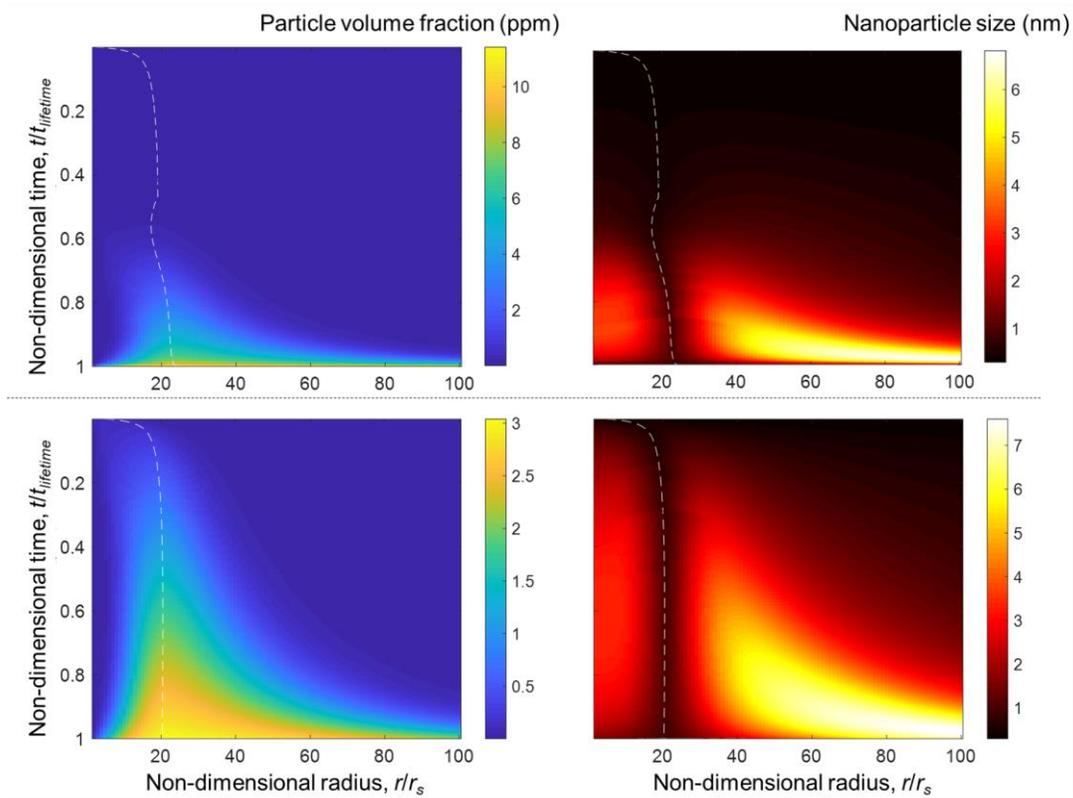

Fig. 5. Evolution of particle volume fraction and nanoparticle size for the liquid precursor with Li-EHA and EtOH (upper panels) and the liquid precursor with TMS and EtOH (lower panels). The white dashed lines indicate the flame sheet position.

Figure 6 further demonstrates the dependence of nanoparticle size on the precursor mass fraction and the initial droplet size inside the flame sheet. The maximum nanoparticle size inside the flame sheet during the droplet lifetime is chosen for analysis. It is found that the precursor mass fraction has a linear relation with the cube of nanoparticle size. The weak dependence of the nanoparticle size on the precursor mass fraction is consistent with the gas-to-particle conversion route in the VAFS [16]. Moreover, as the typical residence time is proportional to the square of droplet size $r_{s0}^2$, the nanoparticle size obeys the relation $d_p \propto r_{s0}^{2/3}$, as demonstrated by the fitting curve.

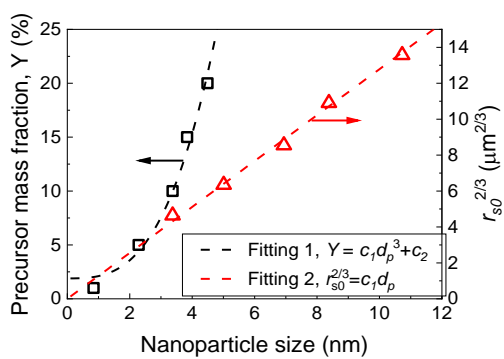

Fig. 6. Dependence of maximum nanoparticle size on precursor concentration and the droplet size.

### 3.3 Droplet-to-particle conversion route

If the precursor decomposes or precipitates in the liquid phase instead of vaporizing, it follows the droplet-to-particle conversion route. Two precursors $Al(NO_3)_3$ and $Zn(NO_3)_2$ mixed with butanol as a solvent are analyzed. The chosen precursors have decomposition temperatures of 423 K and 383 K, which are higher and lower than the solvent boiling point (391 K), respectively. The temperature-time histories of the droplet with different initial radius are demonstrated in Fig. 7. As the nitrates do not vaporize, they concentrate at the outer layer of the droplet, which prevents further vaporization of the solvent. The initial droplet size determines their characteristic evaporation time $t_{evap}$. For smaller droplets, the surface temperature reaches the solvent boiling point $T_{bp,s}$ and the thermal decomposition temperature $T_{dmp}$ earlier.

For $Zn(NO_3)_2$, the $T_{dmp}$ is lower than $T_{bp,s}$. The precursor first decomposes inside the liquid phase, which leads to the solid nuclei. In this condition, for small droplets that can evaporate fast enough, a solid particle would form based on the one-droplet-one-particle rule. On the contrary, if the precursor decomposition occurs first, the liquid precursor transforms into nanoparticles inside the droplet throughout the droplet evaporation process and forms homogeneous fine particles.

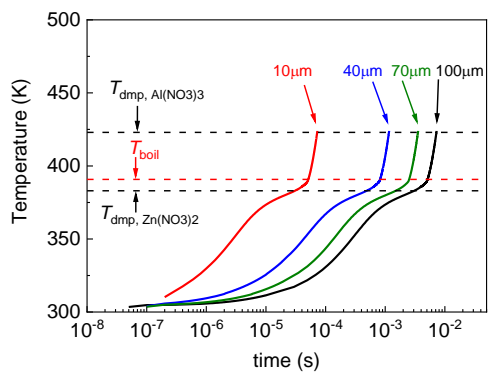

Fig. 7. Temperature-time history of droplet-to-particle conversion.

When $T_{dmp} > T_{bp,s}$, like Al(NO$_3$)$_3$, the droplets behave in the most complicated way as the precursor precipitation is also involved during this process. Based on the simulation of the droplet drying process, it is found that the precipitation occurs when the solvent is heated up to the boiling point, as demonstrated in Fig. 8. As the droplet shrinks, the non-volatile precursor is accumulated at the surface, while the rising rate of the temperature slows down when the solvent reaches the boiling point. The temperature-precursor mass fraction profiles are found to be independent of the droplet size and the nucleation rate but are influenced by the liquid Lewis number. As the droplet shrinkage causes the precursor precipitation, the diffusion-limited species would give rise to a higher local concentration. If the precipitation occurs fast enough, a shell is formed outside the droplet, and micro-explosion occurs, which then induces shell-like products. However, for extreme small droplets that evaporate fast enough, the large droplet evaporation rate will lead to a homogeneous solid particle following the one-droplet-one-particle rule.

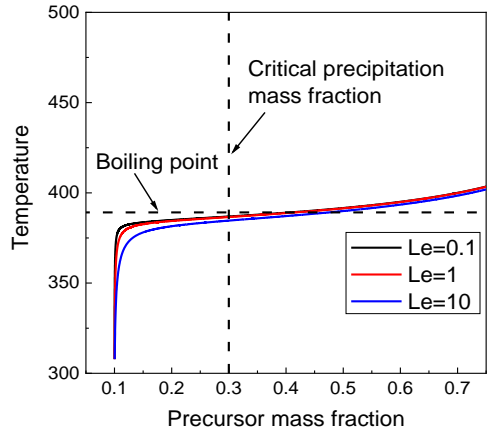

Fig. 8. The temperature and precursor mass fraction profile during the droplet drying.

## 4. Discussion and Conclusion

The results in Section 3 is summarized and demonstrated in a particle formation map in Fig. 9. The final nanoparticle morphology is influenced by both precursor behavior and droplet behavior. The thermal decomposition temperature $T_{dmp,p}$, the precursor boiling point $T_{bp,p}$, the evaporation time $t_{evap}$, and the precursor decomposition time $t_{react}$ together determine the formed nanoparticle morphology. The precursor behavior largely determines the gas-to-particle conversion route or the droplet-to-particle conversion route. If the precursor evaporates before its thermal decomposition, i.e. $T_{dmp,p} > T_{bp,p}$, it will undergo the gas-to-particle conversion route; otherwise, it will follow the particle-to-particle route.

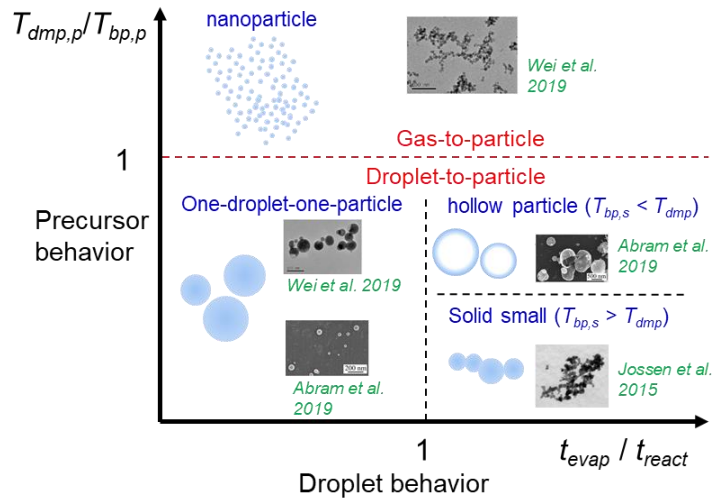

Fig. 9. Particle formation mapping

For the gas-to-particle conversion route, the nanoparticles undergo nucleation, collision, and coagulation like VAFS. The nanoparticle size is weakly determined by the precursor concentration following $d_p \sim c^{1/3}$. In this condition, the external flow field that controls the temperature-time history of the nanoparticle is important. If the boiling points between solvent and precursor differ significantly, like metal-organic precursor, the volatile solvent can be trapped inside the droplet and superheated. In this condition, micro-explosion can easily happen and accelerate the evaporation procedure, as validated in the single-droplet experiments [14].

If the precursor tends to decompose or precipitate in the liquid phase, the droplet-to-particle conversion is also related to the solvent behavior. For ultra-small droplets with $t_{evap}$ smaller than the precursor precipitation time or decomposition time $t_{react}$, the conversion follows the one-droplet-one-particle route [16]. For large droplets with long evaporation time, the competition among precipitation, thermal decomposition, and evaporation determines the final nanoparticle morphology. As the droplet evaporation is limited by the species diffusion, the precursor precipitation occurs near the solvent boiling point. Therefore, If the solvent boiling temperature $T_{bp,s}$ is higher than the precursor thermal decomposition temperature $T_{dmp}$, the precursor reacts in the liquid forming solid small particles [11]. On the contrary, if $T_{bp,s}$ is lower than $T_{dmp}$, the precursor will precipitate at the droplet surface, forming hollow particles.

## Acknowledgment

This work is funded by Alexzander von Humboldt foundation.This work is funded by Alexzander von Humboldt foundation.

**List of Figure Captions**

Fig. 1. The algorithm map of the single droplet model

Fig. 2. Simulated distributions of the mass fractions of fuel, oxidizer, products and inert species, as well as the temperature profile based on the single droplet model. $T_\infty = 500$ K, $Y_{O\infty} = 0.21$, $r_s^2/r_{s0}^2 = 0.99$, $Y_{ethanol} = 0.9$, $Y_{2\text{-EHA}} = 0.1$.

Fig. 3. Temporal variation of surface and center temperatures of a multi-component droplet as the droplet shrinks in a $D^2$-law, which is compared with the PDA measurement data (blue squares) based on Ref. [13]. $T_\infty = 1000$ K, $Y_{O\infty} = 0.21$, $Y_{ethanol} = 0.5$, $Y_{2\text{-EHA}} = 0.5$, $Le_l = 10$.

Fig. 4. Spatial distributions of temperature and species mass fractions of ethanol (blue) and 2-EHA (red) at different moments, labeled as A, B, C in Fig. 3. $T_\infty = 1000$ K, $Y_{O\infty} = 0.21$, $Y_{ethanol} = 0.5$, $Y_{2\text{-EHA}} = 0.5$.

Fig. 5. Evolution of particle volume fraction and nanoparticle size for the liquid precursor with Li-EHA and EtOH (upper panels) and the liquid precursor with TMS and EtOH (lower panels). The white dashed lines indicate the flame sheet position.

Fig. 6. Dependence of maximum nanoparticle size on precursor concentration and the droplet size.

Fig. 7. Temperature-time history of droplet-to-particle conversion.

Fig. 8. The temperature and precursor mass fraction profile during the droplet drying.

Fig. 9. Particle formation mapping.